\documentclass[letterpaper,english,prl, aps, twocolumn, superscriptaddress]{revtex4-1}
\usepackage[T1]{fontenc}
\usepackage[latin9]{inputenc}
\usepackage{mathrsfs}
\usepackage{amsmath}
\usepackage{amssymb}
\usepackage{graphicx}

\makeatletter

\pdfpageheight\paperheight
\pdfpagewidth\paperwidth

\usepackage{xcolor}
\usepackage[colorlinks,citecolor=blue,linkcolor=red,anchorcolor=red,urlcolor=blue]{hyperref}

\makeatother

\usepackage{babel}
\begin{document}
\title{Minimal Energy Cost to Initialize a Quantum Bit with Tolerable Error}
\author{Yu-Han Ma}
\address{Graduate School of China Academy of Engineering Physics, No. 10 Xibeiwang
East Road, Haidian District, Beijing, 100193, China}
\author{Jin-Fu Chen}
\address{Graduate School of China Academy of Engineering Physics, No. 10 Xibeiwang
East Road, Haidian District, Beijing, 100193, China}
\address{Beijing Computational Science Research Center, Beijing 100193, China}
\address{School of Physics, Peking University, Beijing, 100871, China}
\author{C. P. Sun}
\email{suncp@gscaep.ac.cn}

\address{Graduate School of China Academy of Engineering Physics, No. 10 Xibeiwang
East Road, Haidian District, Beijing, 100193, China}
\address{Beijing Computational Science Research Center, Beijing 100193, China}
\author{Hui Dong}
\email{hdong@gscaep.ac.cn}

\address{Graduate School of China Academy of Engineering Physics, No. 10 Xibeiwang
East Road, Haidian District, Beijing, 100193, China}
\date{\today}
\begin{abstract}
Landauer's principle imposes a fundamental limit on the energy cost
to perfectly initialize a classical bit, which is only reached under
the ideal operation with infinite-long time. The question on the cost
in the practical operation for a quantum bit (qubit) has been posted
under the constraint by the finiteness of operation time. We discover
a raise-up of energy cost by $\mathcal{L}^{2}(\epsilon)/\tau$ from
the Landaeur's limit ($k_{B}T\ln2$) for a finite-time $\tau$ initialization
with an error probability $\epsilon$. The thermodynamic length $\mathcal{L}(\epsilon)$
between the states before and after initializing in the parametric
space increases monotonously as the error decreases. For example,
in the constant dissipation coefficient ($\gamma_{0}$) case, the
minimal additional cost is $0.997k_{B}T/(\gamma_{0}\tau)$ for $\epsilon=1\%$
and $1.288k_{B}T/(\gamma_{0}\tau)$ for $\epsilon=0.1\%$. Furthermore,
the optimal protocol to reach the bound of minimal energy cost is
proposed for the qubit initialization realized via a finite-time isothermal
process.
\end{abstract}
\maketitle
\narrowtext

\textit{Introduction. --} Initializing memory is a necessary process
in computation \citep{Landauer1961,Bennett1973,Zurek1989,Bennett1997,Nielsen2000,Parrondo2015}
for further information processing, and inevitably requires an expense
of the energy. Landauer derived a fundamental limit of energy cost
for such process that a minimal $k_{B}T\ln2$ ($k_{B}$ is the Boltzmann
constant) heat will be dissipated to the environment of temperature
$T$ for erasing one bit information \citep{Landauer1961,Berut2012,Jun2014,Peterson2016,Yan2018,dago2021information}.
Such limit is only reached with two idealities, infinite-long operation
time and perfect initialization, which are unfortunately impossible
for practical devices.

The quantum bit (qubit) is the basic unit of quantum computation,
which is believed to surpass the capability of classical computation
\citep{feynman1985quantum,divincenzo1995quantum,Nielsen2000,divincenzo2000physical}.
The short coherence time makes the first ideality of infinite-time
unpractical \citep{shor1995scheme,1995Decoherence}. Fast computation
processes require initializing qubits within the time shorter than
the coherence time if the qubits are reused in these processes. A
finite-time Landauer's principle has been in turn proposed that the
corresponding energy cost is significantly increased with the decrease
of the erasure time \citep{Diana2013,Zulkowski2014,Zulkowski2015a,Proesmans2020,Proesmans2020a,dago2021information,Vu2021}.
And the first ideality is released. When the qubit is initialized
to the desired state, the error of initialization will set up a bound
on the maximum number of the quantum gates. Lower error probability
allows more gate operations. A critical error probability is typically
required in the quantum computation process, e.g., 1\% for error correction
\citep{steane1996error,2011Surface,zhang2020error}. Therefore, the
second ideality is not necessary.

The question is to ask what is the minimal energy cost to initialize
a qubit with tolerable errors within the coherence time? In this Letter,
we tackle this problem by exploiting the geometry framework of quantum
thermodynamic \citep{Ruppeiner1979,Ruppeiner1995} to study the finite-time
information erasure in a qubit. The erasure here is realized by driving
the qubit in a thermal bath. When the driving is not applied infinite-slowly,
the additional work required in the erasure process is quantified
by the thermodynamic length $\mathcal{L}$ \citep{Ruppeiner1979,Salamon1983,Gilmore1984,Diosi1996,Crooks2007,Sivak2012,Chen2021},
which depends on the error probability $\epsilon$. We discover an
analytical trade-off relation among the energy cost, erasure time,
and error probability with the asymptotic behavior of $\mathcal{L}$.
For applications, we demonstrate the error-probability-dependent optimal
erasure protocol to achieve the minimal energy cost for qubit initialization.

\begin{figure}
\centering{}\includegraphics[width=8.5cm]{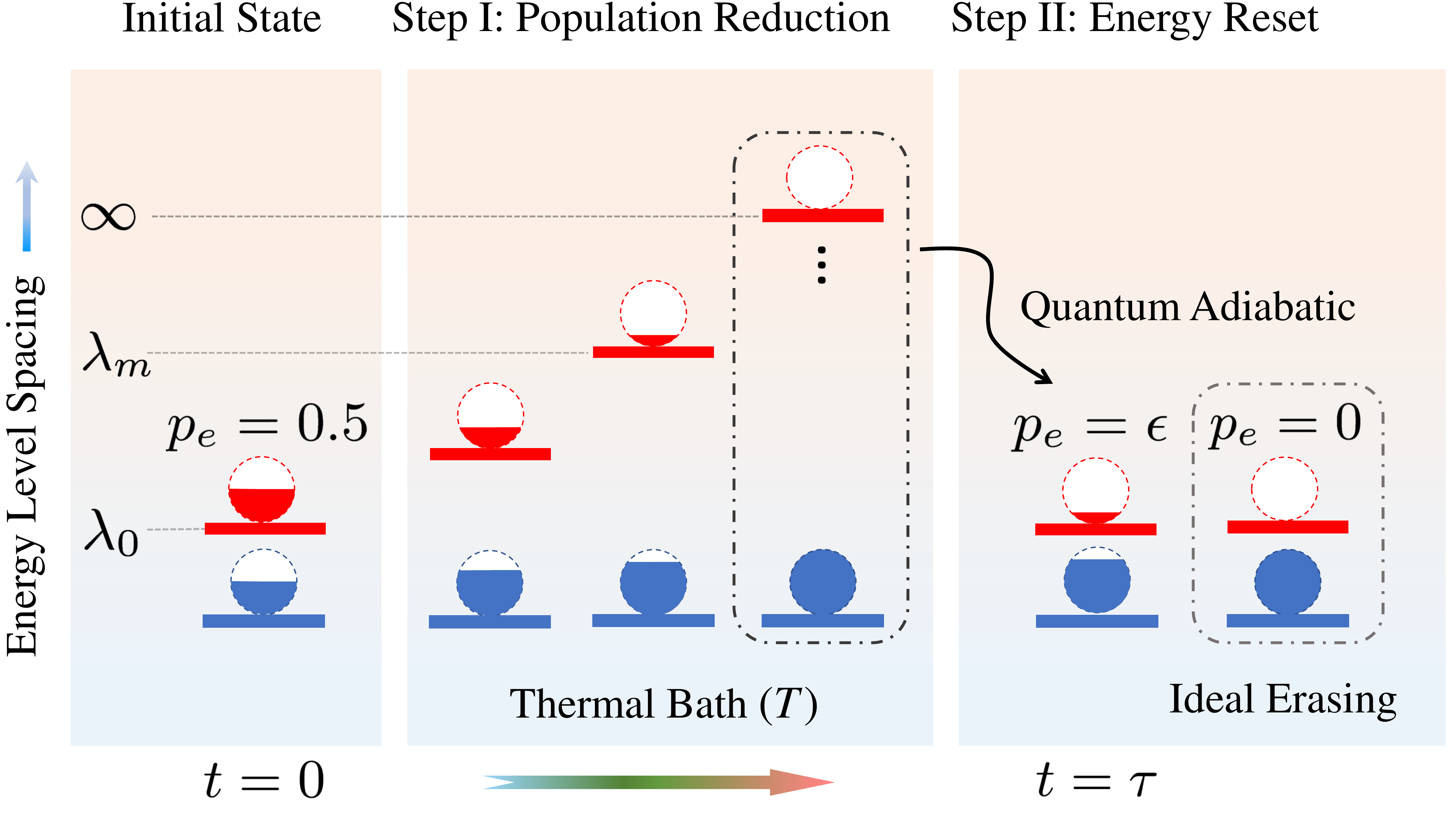}\caption{\label{fig:Finite-time-information-erasing}Schematic of finite-time
information erasure for a quantum bit (qubit). The ground state and
excited state of the qubit represent the logical states \textquotedbl$0$\textquotedbl{}
and \textquotedbl$1$\textquotedbl , respectively. The energy level
spacing of the qubit $\lambda$ is firstly tuned from $\lambda_{0}=0$
to $\lambda_{f}=\lambda_{m}$ when it is in contact with a thermal
bath of temperature $T$. Then the energy level spacing of the qubit
is tuned back to $0$ quantum adiabatically. In the ideal erasure
process with $\lambda_{m}\rightarrow\infty$, the qubit is completely
erased to the logical state \textquotedbl$0$\textquotedbl{} with
the population in the logical state \textquotedbl$1$\textquotedbl{}
$p_{e}=0$.}
\end{figure}

\textit{Work cost for initializing a qubit. --} The qubit is physically
modeled as a two-level system, and the information encoded by the
logical state \textquotedbl$0$\textquotedbl{} (\textquotedbl$1$\textquotedbl )
is represented by the ground state $|g\rangle$ (excited state $|e\rangle$)
of the two-level system. Before the erasure process, we assume that
the qubit contains one bit of information and stays at the maximum
mixed state $\rho_{i}=(|e\rangle\langle e|+|g\rangle\langle g|)/2$.
In the ideal erasure, the qubit is restored into the logical state
\textquotedbl$0$\textquotedbl{} perfectly, namely $\rho_{f}=|g\rangle\langle g|$.
However, in a practical process, the qubit is generally erased to
$\rho_{f}=\epsilon|e\rangle\langle e|+(1-\epsilon)|g\rangle\langle g|$
with an error probability $\epsilon$\textbf{.}

To implement the initialization, the energy level spacing of the two-level
system is tuned by a control parameter $\lambda(t)$ under the Hamiltonian
$H(t)=\lambda(t)\sigma_{z}/2$ with the Pauli matrix $\sigma_{z}=|e\rangle\langle e|-|g\rangle\langle g|$.
The Planck's constant is taken as $\hbar=1$ hereafter for brevity.
As illustrated in Fig. \ref{fig:Finite-time-information-erasing},
the whole information erasure process is designed with two steps as
follows. \textbf{(i)} \textbf{Population reduction} by increasing
the energy level spacing $\lambda$ from $\lambda_{0}=0$ to $\lambda_{m}$.
The qubit is coupled to a thermal bath of inverse temperature $\beta=1/(k_{B}T)$.
\textbf{(ii)} \textbf{Energy reset} by quantum adiabatically decreasing
the energy level spacing of the qubit back to $\lambda_{0}$ with
no thermal bath. The first step is designed to reduce the population
in the excited state, and the second step aims to restore the system's
Hamiltonian for future operation.

The total work performed in the erasure process with duration $\tau$
is $W(\tau)=\int_{0}^{\tau}\dot{W}dt$ with the erasure power $\dot{W}\equiv\mathrm{tr}(\rho\dot{H})$
\citep{Alicki1979,Quan2007}, which is explicitly

\begin{equation}
\dot{W}=\frac{\dot{\lambda}(t)}{2}\left[2p_{e}(t)-1\right].\label{eq:Wdot}
\end{equation}
The probability for the qubit on the excited state $p_{e}(t)$ is
governed by the master equation \citep{Breuer2007} $\dot{p}_{e}=\mathscr{L}(p_{e})$.
For the qubit in contact with a bosonic heat bath, $\mathscr{L}(p_{e})$
reads \citep{Breuer2007}, 
\begin{equation}
\mathscr{L}(p_{e})=\gamma\left\{ n(\lambda)-\left[2n(\lambda)+1\right]p_{e}\right\} ,\label{eq:master equation}
\end{equation}
where $n(\lambda)=1/(e^{\beta\lambda}-1)$ is the average particle
number of the bath mode with energy $\lambda$, and the dissipation
coefficient $\gamma=\gamma(\lambda)$ is determined by the bath spectral
\citep{Breuer2007}.

To evaluate the finite-time effect, we define the irreversible work
$W_{\mathrm{ir}}(\tau)\equiv W(\tau)-\Delta F=\int_{0}^{\tau}\dot{W}_{\mathrm{ir}}(t)dt$,
where $\Delta F=\beta^{-1}\left[\ln2-S(\epsilon)\right]$ \citep{SM}
is the free energy change between the states of the qubit before and
after initialization ($S(\epsilon)=-\epsilon\ln\epsilon-(1-\epsilon)\ln(1-\epsilon)$
is the Shannon entropy of the final state). The irreversible power
\citep{Sivak2012,Chen2021} is

\begin{equation}
\dot{W}_{\mathrm{ir}}=\dot{\lambda}(t)\left\{ p_{e}(t)-p_{e}^{\mathrm{eq}}\left[\lambda(t)\right]\right\} ,\label{eq:Wir-general}
\end{equation}
where $p_{e}^{\mathrm{eq}}[\lambda(t)]=e^{-\beta\lambda(t)}/[1+e^{-\beta\lambda(t)}]$
is the excited-state population in the instantaneous thermal equilibrium
distribution. Finding the minimal energy cost for erasing the information
stored a qubit is now converted to deriving the lower bound of the
irreversible work.

In the slow-driving regime with $\gamma\tau\gg1$, the irreversible
power to the first order of $1/(\gamma\tau)$ is \citep{SM}

\begin{equation}
\dot{W}_{\mathrm{ir}}=\beta\gamma^{-1}\frac{\left(1-e^{-\beta\lambda}\right)e^{-\beta\lambda}}{\left(1+e^{-\beta\lambda}\right)^{3}}\dot{\lambda}^{2}.\label{eq:Wir-dot}
\end{equation}
The typical form of the dissipation coefficient $\gamma=\gamma_{0}\lambda^{\alpha}$
will be used in the following discussion, where $\gamma_{0}$ is a
constant. And $\alpha\in\left[0,1\right)$, $\alpha=1$, and $\alpha>1$
correspond to sub-Ohmic, Ohmic, and super-Ohmic spectral, respectively
\citep{leggett1987dynamics,Breuer2007,Scandi2018}.

\textit{Energy-time-error trade-off. }-- The irreversible work \textit{$W_{\mathrm{ir}}$}
is bounded from below by $W_{\mathrm{ir}}\geq\mathcal{L}^{2}/\tau$
with the thermodynamic length $\mathcal{L}\equiv\int_{0}^{\tau}\sqrt{\dot{W}_{\mathrm{ir}}}dt$
\citep{Ruppeiner1979,Salamon1983,Gilmore1984,Diosi1996,Crooks2007,Sivak2012,Brandner2020,Abiuso2020a,Chen2021},
which is the geometric distance in the parametric space and independent
of the specific erasure protocol of $\lambda(t)$. The thermodynamic
length for the erasure process is obtained explicitly as $\mathcal{L}(\epsilon)=\sqrt{\beta^{(\alpha-1)}\gamma_{0}^{-1}}f_{\alpha}(\epsilon)$
with the dimensionless function \citep{SM}

\begin{equation}
f_{\alpha}(\epsilon)\equiv\int_{0}^{\ln\left(\epsilon^{-1}-1\right)}\sqrt{\frac{\left(1-e^{-x}\right)e^{-x}}{x^{\alpha}\left(1+e^{-x}\right)^{3}}}dx.\label{eq:f}
\end{equation}
The upper limit of the integral in Eq. (\ref{eq:f}) reflects the
dependence of $\mathcal{L}(\epsilon)$ on the error probability $\epsilon$.
The precise lower bound for irreversible work is denoted as $W_{\mathrm{ir}}^{\mathrm{min}}(\epsilon)\equiv\mathcal{L}^{2}(\epsilon)/\tau$.
Particularly, in the constant dissipation coefficient case ($\alpha=1$),
$W_{\mathrm{ir}}^{\mathrm{min}}=0.997k_{B}T/(\gamma_{0}\tau)$ for
$\epsilon=1\%$ and $W_{\mathrm{ir}}^{\mathrm{min}}=1.288k_{B}T/(\gamma_{0}\tau)$
for $\epsilon=0.1\%$.

Utilizing the asymptotic behavior of the thermodynamic length $\left|\mathcal{L}(\epsilon)-\mathcal{L}(0)\right|\propto\sqrt{\epsilon\ln^{-\alpha}\left(\epsilon^{-1}\right)}$
, we obtain the main result of this Letter \citep{SM}

\begin{equation}
\frac{W_{\mathrm{ir}}\tau}{\mathcal{L}^{2}(0)}+\mu_{\alpha}\sqrt{\epsilon\ln^{-\alpha}\left(\epsilon^{-1}\right)}\geq1.\label{eq:trade-off}
\end{equation}
Here, $\mathcal{L}(0)$ is the thermodynamic length for the perfect
erasing ($\epsilon=0$), and $\mu_{\alpha}\equiv4/f_{\alpha}(0)$
is a dimensionless constant determined solely by the bath spectral.
This inequality quantitatively reveals a trade-off relation among
irreversible work $W_{\mathrm{ir}}$, erasure time $\tau$, and error
probability $\epsilon$. For the special case of the perfect erasing
($\epsilon=0$), such a trade-off recovers the result $W_{\mathrm{ir}}\geq\mathcal{L}^{2}(0)/\tau$
obtained in the recent studies on finite-time Landauer's principle
\citep{Berut2012,Zulkowski2014,Zulkowski2015a,Proesmans2020,Proesmans2020a,Miller2020,Vu2021,Dago2021}.

\begin{figure}
\begin{raggedright}
\includegraphics[width=8.5cm]{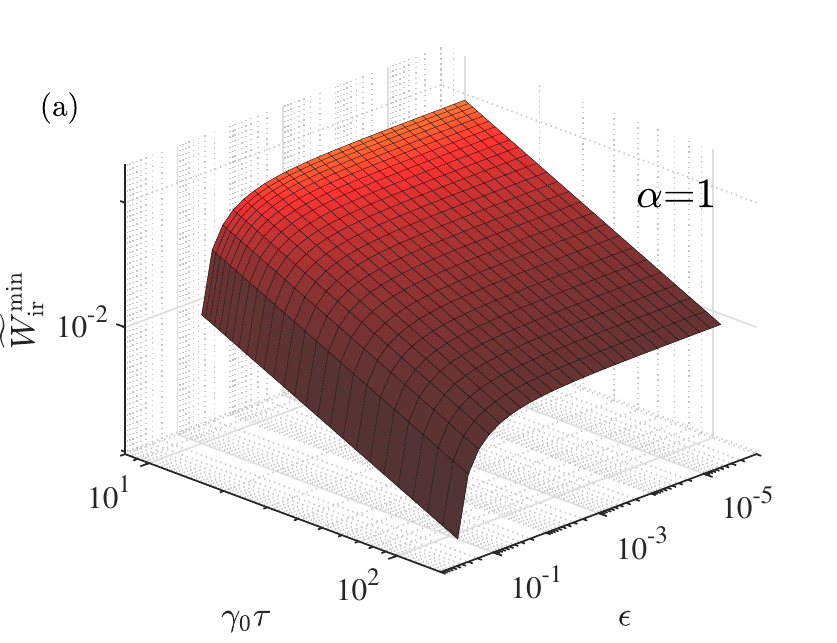}
\par\end{raggedright}
\begin{raggedright}
\includegraphics[width=8.5cm]{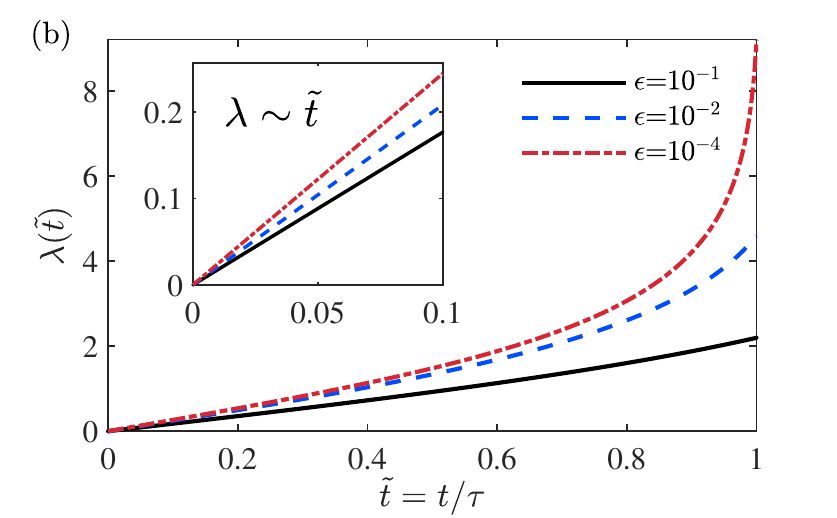}
\par\end{raggedright}
\centering{}\caption{\label{fig:Analytical-trade-off-between} \textbf{(a)} Analytical
trade-off among irreversible work, erasure time $\tau$ and error
probability $\epsilon$ in the Ohmic spectral case ($\alpha=1$).
The surface is plotted with the analytical lower bound ($\widetilde{W}_{\mathrm{ir}}^{\mathrm{min}}$)
of Eq. (\ref{eq:trade-off}) with $f_{1}(0)=0.9433$ \citep{SM}.
\textbf{(b)} Optimal protocol of $\lambda(\widetilde{t})$ in the
Ohmic spectral case ($\alpha=1$) with different error probabilities
$\epsilon$. The black solid curve, blue dashed curve, and red dash-dotted
curve represent the optimal protocol for $\epsilon=10^{-1}$, $\epsilon=10^{-2}$,
and $\epsilon=10^{-4}$, respectively. The scaling of the optimal
protocol in the initial stage $\lambda(\widetilde{t}\ll1)\sim\widetilde{t}$
is illustrated in the inset figure. The parameters $\beta=1$ and
$\gamma_{0}=1$ are used to plot this figure.}
\end{figure}

We plot the analytical lower bound $\widetilde{W}_{\mathrm{ir}}^{\mathrm{min}}\equiv[1-\mu_{\alpha}\sqrt{\epsilon\ln^{-\alpha}(\epsilon^{-1})}]\mathcal{L}^{2}(0)/\tau$
of Eq. (\ref{eq:trade-off}) as the surface in Fig. \ref{fig:Analytical-trade-off-between}
(a) for the Ohmic spectral ($\alpha=1$) case. In the simulation,
we always use the following parameters $\gamma_{0}=1$ and $\beta=1$.
The monotonicity of the surface indicates that more extra energy cost
is required to accomplish the erasure with higher accuracy and shorter
operation time. The trade-offs in the cases with $\alpha=0,2$ are
illustrated in Supplementary Materials (SM) \citep{SM}.

\textit{The optimal erasure protocol.} -- The optimal erasure protocol
$\lambda(t)$ applied to initialize the qubit for minimal work satisfies
$\sqrt{\dot{W}_{\mathrm{ir}}}=\mathcal{L}(\epsilon)/\tau$, namely,

\begin{equation}
\frac{d\lambda}{d\widetilde{t}}=\mathcal{L}(\epsilon)\left[\frac{\beta\left(1-e^{-\beta\lambda}\right)e^{-\beta\lambda}}{\gamma_{0}\lambda^{\alpha}\left(1+e^{-\beta\lambda}\right)^{3}}\right]^{-\frac{1}{2}},\label{eq:equationforlambda}
\end{equation}
with the dimensionless time $\tilde{t}\equiv t/\tau$. The exact optimal
protocol is numerically obtained by solving the above differential
equation with the boundary conditions $\lambda(\tilde{t}=0)=0$ and
$\lambda(\tilde{t}=1)=\lambda_{m}$. In Fig. \ref{fig:Analytical-trade-off-between}
(b), we illustrate the optimal protocols for cases with $\alpha=1$
for $\epsilon=10^{-1}$ (black solid curve), $\epsilon=10^{-2}$ (blue
dashed curve), and $\epsilon=10^{-4}$ (red dash-dotted curve), respectively.
The optimal protocols in the cases with $\alpha=0,2$ are demonstrated
in SM \citep{SM}. In the initial stage ($t/\tau\ll1$) of the erasure
process, we find the optimal protocols satisfy a universal scaling
$\lambda(\widetilde{t}\ll1)\sim\widetilde{t}^{2/(3-\alpha)}$ by solving
Eq. (\ref{eq:equationforlambda}) approximately \citep{SM}. The inset
figure shows that the optimal protocol scales as $\lambda\sim\widetilde{t}$
in the present case ($\alpha=1$). Additionally, we stress that the
solution of Eq. (\ref{eq:equationforlambda}) with respect to $\widetilde{t}$
is independent of the erasure time. Therefore, for given erasure processes
(fixed $\alpha,\epsilon$) with different duration $\tau$, the required
optimal protocol $\lambda(t)$ can be directly obtained by performing
variable substitution $\tilde{t}\rightarrow t/\tau$ on the fixed
optimal protocol (with respect to $\widetilde{t}$).

\begin{figure}
\begin{centering}
\includegraphics[width=8.5cm]{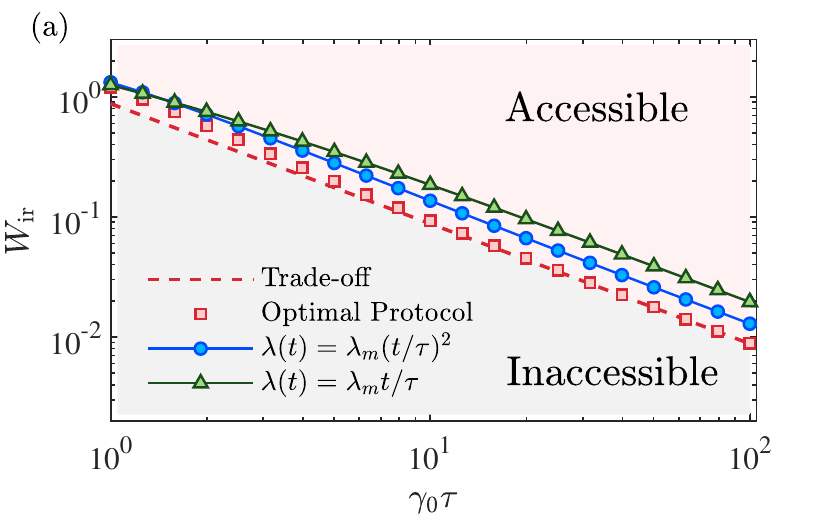}
\par\end{centering}
\begin{centering}
\includegraphics[width=8.5cm]{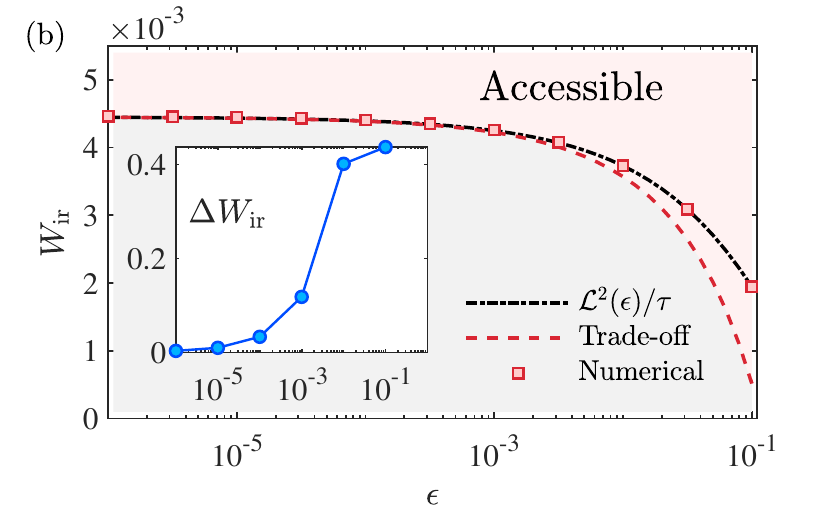}
\par\end{centering}
\centering{}\caption{\label{fig:validation} \textbf{(a)} Irreversible work as a function
of the erasure time $\tau$ in the Ohmic spectral case ($\alpha=1$),
where $\epsilon=10^{-4}$ is fixed. The red squares represent the
exact minimal irreversible work associated with the optimal erasure
protocol. The blue dot curve (green triangle curve) is obtained numerically
with the erasure protocol chosen as $\lambda(t)=\lambda_{m}(t/\tau)^{2}$
($\lambda(t)=\lambda_{m}t/\tau$), where $\lambda_{m}=\beta^{-1}\ln(\epsilon^{-1}-1)$.
The red dashed curve is plotted with the analytical lower bound $\widetilde{W}_{\mathrm{ir}}^{\mathrm{min}}$.
\textbf{(b)} Minimal irreversible work as a function of error probability
$\epsilon$ in the case with $\alpha=1$, where $\tau=200$ is fixed.
The red squares are obtained numerically with the optimal erasure
protocol, while the black dash-dotted curve and red dashed curve represent
the precise lower bound $W_{\mathrm{ir}}^{\mathrm{min}}=\mathcal{L}^{2}(\epsilon)/\tau$
and the analytical lower bound $\widetilde{W}_{\mathrm{ir}}^{\mathrm{min}}$,
respectively. The additional work $\Delta W_{\mathrm{ir}}(\epsilon)$
required to reduce the error probability $\epsilon$ by an order of
magnitude is plotted in the inset figure. In this figure, we use $\beta=1$
and $\gamma_{0}=1$.}
\end{figure}

\textit{Numerical validations of the trade-off.} -- We solve the
exact minimal irreversible work numerically from Eqs. (\ref{eq:master equation})
and (\ref{eq:Wir-general}) to validate the analytical trade-off in
Eq. (\ref{eq:trade-off}). The irreversible work is illustrated in
Fig. \ref{fig:validation}(a) as a function of the erasure time with
$\epsilon=10^{-4}$. The red squares represent the numerical results
corresponding to the optimal erasure protocol, and the analytical
lower bound $\widetilde{W}_{\mathrm{ir}}^{\mathrm{min}}$ is plotted
with red dash-dotted curve. The analytical trade-off, exhibiting the
typical $1/\tau$-scaling of irreversibility \citep{Salamon1980a,Salamon1983,sekimoto1997complementarity,Berut2012,Ma2018,Ma2019,dago2021information},
is in good agreement with the numerical results in the long-time regime
of $\gamma_{0}\tau\gg1$. In the short-time regime (beyond the slow-driving
regime), the higher order terms of $1/(\gamma_{0}\tau)$ in the expansion
of $\dot{W}_{\mathrm{ir}}$ can not be ignored anymore \citep{SM},
and thus the minimal irreversible work deviates from the $1/\tau$-scaling
\citep{Ma2018,yhmaoptimalcontrol,Ma2019}. To demonstrate the dependence
of irreversible work on the erasure protocol, the exact numerical
irreversible work related to two erasure protocols, $\lambda(t)=\lambda_{m}(t/\tau)^{2}$
and $\lambda(t)=\lambda_{m}t/\tau$ ($\lambda_{m}=\beta^{-1}\ln(\epsilon^{-1}-1)$),
are illustrated with the blue dot curve and green triangle curve,
respectively. The irreversible work corresponding to these two protocols
are larger in comparison with the minimal irreversible work achieved
with the optimal protocol (red squares). Since the irreversible work
can not be less than the analytical lower bound $\widetilde{W}_{\mathrm{ir}}^{\mathrm{min}}$
with any erasure protocol, we denote the light gray area below the
red dash-dotted curve as the inaccessible regime of the irreversible
work. In this sense, the light red area above the red dash-dotted
curve is accessible.

The exact minimal irreversible work as a function of the error probability
is marked with the red squares in Fig. \ref{fig:validation} (b) for
fixed duration $\tau=200$. The irreversible work increases for lower
the error probability. Similar to Fig. \ref{fig:validation} (a),
the two areas separated by the red dashed curve (analytical lower
bound of Eq. (\ref{eq:trade-off})) represent the accessible and inaccessible
regions. The black dash-dotted curve represents the precise bound
of the irreversible work $W_{\mathrm{ir}}^{\mathrm{min}}$ characterized
by the exact thermodynamic length, which agrees well with the exact
numerical results in the whole plotted range of $\epsilon$. The fact
that the red dashed curve ($\widetilde{W}_{\mathrm{ir}}^{\mathrm{min}}$)
approaches to the black dash-dotted curve ($W_{\mathrm{ir}}^{\mathrm{min}}$)
in the low error probability regime ($\epsilon\ll1$) is consistent
with the approximation condition used to obtain Eq. (\ref{eq:trade-off}).
In addition, we introduce the following normalized quantity $\Delta W_{\mathrm{ir}}(\epsilon)\equiv[W_{\mathrm{ir}}^{\mathrm{min}}(\epsilon)-W_{\mathrm{ir}}^{\mathrm{min}}(10\epsilon)]/W_{\mathrm{ir}}^{\mathrm{min}}(0)$
to evaluate the additional work required to reduce the error probability
$\epsilon$ by an order of magnitude. As demonstrated in the inset
figure of Fig. \ref{fig:validation} (b), $\Delta W_{\mathrm{ir}}$
decreases rapidly with the error probability. We remark that it typically
requires less additional work to reduce the error probability in the
low-$\epsilon$ regime, noticing the plateau at $\epsilon\leq10^{-4}$
region.

The numerical results confirm that the analytical trade-off (red dashed
curve) approaches the precise lower bound (black dash-dotted curve)
with the error probability $\epsilon\ll1$. Beyond the regime $\epsilon\ll1$
or $\gamma\tau\gg1$, one can observe from Fig. \ref{fig:Analytical-trade-off-between}
(a) and (b) that all the exact minimal irreversible work (red squares)
located above the red dashed curve. This implies that our analytical
trade-off (\ref{eq:trade-off}) may have a wider applicable scope
beyond the slow-driving regime.

\textit{Conclusions and discussions. --} In this Letter, we studied
the finite-time information erasure in a qubit with tolerable errors.
A universal trade-off among irreversible work, erasure time, and error
probability is obtained for non-equilibrium erasure processes characterized
by the thermodynamic length. This trade-off relation reveals that
reducing the erasure time and error probability require additional
energy cost. For practical purposes, we found the optimal erasure
protocol associated with the minimal work cost to initialize a qubit.
This study paves the way for analyzing the finite-time information
processing with the thermodynamic geometry approach, and shall bring
new insights to the practical optimization of gate control in quantum
computation. As possible extensions, the influences of different bath
spectral, quantum coherence of the qubit \citep{Su2018,Miller2020,Vu2021},
and fast-driving of the erasure process \citep{Dago2021,Chen2021,Fast-driving,Fast-drivingMa}
on the trade-off obtained in the current work can be taken into future
consideration.

\textit{Acknowledgments.} -- This work is supported by the National
Natural Science Foundation of China (NSFC) (Grants No. 11534002, No.
11875049, No. U1730449, No. U1530401, and No. U1930403), the National
Basic Research Program of China (Grants No. 2016YFA0301201), and the
China Postdoctoral Science Foundation (Grant No. BX2021030).

\textit{Note added:} After completion of this work, we became aware
of a recent related work by Zhen et al. \citep{Zhen2021}

\bibliographystyle{apsrev}
\bibliography{mainref}

\end{document}

\title{\textit{\Large{}Supplementary Materials for}{\Large{} \textquotedbl Minimal
Energy Cost to Initialize a Quantum Bit with Tolerable Error?\textquotedbl}}
\author{Yu-Han Ma}
\address{Graduate School of China Academy of Engineering Physics, No. 10 Xibeiwang
East Road, Haidian District, Beijing, 100193, China}
\author{Jin-Fu Chen}
\address{Graduate School of China Academy of Engineering Physics, No. 10 Xibeiwang
East Road, Haidian District, Beijing, 100193, China}
\address{Beijing Computational Science Research Center, Beijing 100193, China}
\address{School of Physics, Peking University, Beijing, 100871, China}
\author{C. P. Sun}
\email{suncp@gscaep.ac.cn}

\address{Graduate School of China Academy of Engineering Physics, No. 10 Xibeiwang
East Road, Haidian District, Beijing, 100193, China}
\address{Beijing Computational Science Research Center, Beijing 100193, China}
\author{Hui Dong}
\email{hdong@gscaep.ac.cn}

\address{Graduate School of China Academy of Engineering Physics, No. 10 Xibeiwang
East Road, Haidian District, Beijing, 100193, China}
\date{\today}
\maketitle
\begin{widetext}
In Sec. \ref{sec:Quasi-static-erasing-work} of the Supplementary
Materials, we show the detailed calculations of the free energy change
$\Delta F$ and the irreversible power $\dot{W}_{\mathrm{ir}}$ in
the erasure process. The asymptotic expression of the thermodynamic
length is derived in Sec. \ref{sec:Thermodynamic-length}, and the
trade-offs in the cases with $\alpha=0,2$ are demonstrated. Finally,
in Sec. \ref{sec:Optimal-control-for}, we show the optimal erasure
protocols in the cases with $\alpha=0,2$ and present the analytical
discussion on the optimal protocol in the initial stage.
\end{widetext}

\section{\label{sec:Quasi-static-erasing-work}Free energy change and irreversible
power}

\subsection{Free energy change}
\begin{widetext}
The free energy change $\Delta F$ between the states of the qubit
before and after initializing is equal to the work done in the quasi-static
erasure process. In the quasi-static population reduction process,
the qubit always satisfies the thermal equilibrium distribution $p_{e}^{\mathrm{eq}}(\lambda)=e^{-\beta\lambda}/(1+e^{-\beta\lambda})$.
When the energy level spacing of the qubit is tuned from $\lambda_{0}\rightarrow\lambda_{m}$,
the corresponding quasi-static work $W_{\mathrm{qs}}^{(1)}\equiv\int_{\lambda_{0}}^{\lambda_{m}}\left[p_{e}^{\mathrm{eq}}(\lambda)-1/2\right]d\lambda$
is obtained as

\begin{equation}
W_{\mathrm{qs}}^{(1)}=\beta^{-1}\ln\left(\frac{1+e^{-\beta\lambda_{0}}}{1+e^{-\beta\lambda_{m}}}\right)-\frac{1}{2}\left(\lambda_{m}-\lambda_{0}\right).
\end{equation}
Then the energy level spacing of the qubit is tuned back to $\lambda_{0}$
adiabatically in the energy reset process, and the work done $W_{\mathrm{qs}}^{(2)}\equiv\int_{\lambda_{m}}^{\lambda_{0}}\left[p_{e}^{\mathrm{eq}}(\lambda_{m})-1/2\right]d\lambda$
is

\begin{equation}
W_{\mathrm{qs}}^{(2)}=\frac{e^{-\beta\lambda_{m}}}{1+e^{-\beta\lambda_{m}}}\left(\lambda_{0}-\lambda_{m}\right)+\frac{1}{2}\left(\lambda_{m}-\lambda_{0}\right).
\end{equation}
The total quasi-static work done in these two processes $W_{\mathrm{qs}}\equiv W_{\mathrm{qs}}^{(1)}+W_{\mathrm{qs}}^{(2)}$
is obtained as

\begin{equation}
W_{\mathrm{qs}}=\Delta F=\beta^{-1}\left[\ln2-S(\epsilon)\right],
\end{equation}
Here, we have chosen $\lambda_{0}=0,$ and defined $\epsilon\equiv e^{-\beta\lambda_{m}}/(1+e^{-\beta\lambda_{m}})$
as the error probability. $S(\epsilon)=-\epsilon\ln\epsilon-(1-\epsilon)\ln(1-\epsilon)$
is the Shannon entropy of the final state, and $\beta^{-1}\ln2$ is
the work cost in the ideal erasure process ($\epsilon=0$) as stated
by \textit{Landauer's principle} \citep{Landauer1961}. Since the
entropy $S(\epsilon)$ is an increasing function for $0\leq\epsilon\leq1/2$,
more work is required to achieve lower error probability (smaller
$\epsilon$).
\end{widetext}

\subsection{Irreversible power}
\begin{widetext}
When the qubit is in contact with the bosonic heat bath, the evolution
of the excited-state population is governed by the master equation
\citep{Breuer2007}
\begin{equation}
\dot{p}_{e}=\gamma(\lambda)\left\{ n(\lambda)-\left[2n(\lambda)+1\right]p_{e}\right\} ,\label{eq:master equation}
\end{equation}
where $n(\lambda)=1/(e^{\beta\lambda}-1)$ is the average particle
number of the bath mode with energy $\lambda$, and the dissipation
coefficient $\gamma(\lambda)$ is determined by the bath spectral
\citep{leggett1987dynamics}. With the equilibrium distribution $p_{e}^{\mathrm{eq}}(\lambda)=e^{-\beta\lambda}/(1+e^{-\beta\lambda})$,
Eq. (\ref{eq:master equation}) is rewritten as

\begin{equation}
\left[1+\frac{1-2p_{e}^{\mathrm{eq}}(\lambda)}{\gamma(\lambda)\tau}\frac{d}{d\widetilde{t}}\right]p_{e}=p_{e}^{\mathrm{eq}}(\lambda),
\end{equation}
with the dimensionless time $\widetilde{t}\equiv t/\tau$ . The series
expansion solution of this equation with respect to $\gamma(\lambda)\tau$
follows as

\begin{align}
p_{e} & =\left[1+\frac{1-2p_{e}^{\mathrm{eq}}(\lambda)}{\gamma(\lambda)\tau}\frac{d}{d\widetilde{t}}\right]^{-1}p_{e}^{\mathrm{eq}}(\lambda)\\
 & =\sum_{n=0}\left[-\frac{1-2p_{e}^{\mathrm{eq}}(\lambda)}{\gamma(\lambda)\tau}\frac{d}{d\widetilde{t}}\right]^{n}p_{e}^{\mathrm{eq}}(\lambda).
\end{align}
In the slow-driving regime with $\gamma(\lambda)\tau\gg1$, we obtain
the excited population $p_{e}(t)$ to the first order of $1/\left[\gamma(\lambda)\tau\right]$
as

\begin{equation}
p_{e}\approx p_{e}^{\mathrm{eq}}(\lambda)-\frac{1-2p_{e}^{\mathrm{eq}}(\lambda)}{\gamma(\lambda)}\frac{\partial p_{e}^{\mathrm{eq}}}{\partial\lambda}\dot{\lambda}.\label{eq:pe-peq-1}
\end{equation}
The irreversible power $\dot{W}_{\mathrm{ir}}=\dot{\lambda}(t)(p_{e}-p_{e}^{\mathrm{eq}})$
is thus explicitly obtained as

\begin{align}
\dot{W}_{\mathrm{ir}} & =-\frac{1-2p_{e}^{\mathrm{eq}}(\lambda)}{\gamma(\lambda)}\frac{\partial p_{e}^{\mathrm{eq}}}{\partial\lambda}\dot{\lambda}^{2}\label{eq:}\\
 & =\frac{\beta}{\gamma(\lambda)}\frac{\left(1-e^{-\beta\lambda}\right)e^{-\beta\lambda}}{\left(1+e^{-\beta\lambda}\right)^{3}}\dot{\lambda}^{2}.\label{eq:Wir-lambda}
\end{align}
\end{widetext}

\section{\label{sec:Thermodynamic-length}Asymptotic expression of the thermodynamic
length}
\begin{widetext}
In this section, we discuss the asymptotic expression of the thermodynamic
length $\mathcal{L}(\epsilon)$ with respect to $\epsilon$, and obtain
approximately an analytical trade-off among irreversible work, erasure
time, and error probability.

In the slow-driving regime, by using Eq. (\ref{eq:}) and the typical
dissipation coefficient $\gamma(\lambda)=\gamma_{0}\lambda^{\alpha}$
\citep{Breuer2007}, the thermodynamic length $\mathcal{L}\equiv\int_{0}^{\tau}\sqrt{\dot{W}_{\mathrm{ir}}}dt$
\citep{Sivak2012} of the population reduction process is obtained
in terms of $\epsilon$ as

\begin{align}
\mathcal{L}(\epsilon) & =\sqrt{\beta^{\left(\alpha-1\right)}\gamma_{0}^{-1}}\int_{0}^{\ln\left(\epsilon^{-1}-1\right)}\sqrt{\frac{\left(1-e^{-x}\right)e^{-x}}{x^{\alpha}\left(1+e^{-x}\right)^{3}}}dx\label{eq:TL}\\
 & \equiv\sqrt{\beta^{\left(\alpha-1\right)}\gamma_{0}^{-1}}f_{\alpha}(\epsilon).
\end{align}
For the perfect erasure $\epsilon=0$, one has $\mathcal{L}(0)=\sqrt{\beta^{\left(\alpha-1\right)}\gamma_{0}^{-1}}f_{\alpha}(0)$
with

\begin{equation}
f_{\alpha}(0)=\int_{0}^{\infty}\sqrt{\frac{\left(1-e^{-x}\right)e^{-x}}{x^{\alpha}\left(1+e^{-x}\right)^{3}}}dx=\begin{cases}
1.1981 & \alpha=0\\
0.9433 & \alpha=1\\
1.0914 & \alpha=2
\end{cases}.
\end{equation}
For general erasure processes, $f_{\alpha}(\epsilon)$ can be re-expressed
in term of $f_{\alpha}(0)$ as $f_{\alpha}(\epsilon)=f_{\alpha}(0)-I_{\alpha}(\epsilon)$
with

\begin{equation}
I_{\alpha}(\epsilon)\equiv\int_{\ln\left(\epsilon^{-1}-1\right)}^{\infty}\sqrt{\frac{\left(1-e^{-x}\right)e^{-x}}{x^{\alpha}\left(1+e^{-x}\right)^{3}}}dx.\label{eq:f-a-e-1}
\end{equation}
 We further use a new integral variable $y\equiv e^{-x}$ to rewrite
Eq. (\ref{eq:f-a-e-1}) as
\begin{equation}
I_{\alpha}(\epsilon)=\int_{0}^{\epsilon}\left(-\ln y\right)^{-\alpha/2}\sqrt{\frac{\left(1-y\right)}{y\left(1+y\right)^{3}}}dy,\label{eq:I1}
\end{equation}
which is expanded into series form with respect to $y$ as

\begin{equation}
I_{\alpha}(\epsilon)=\int_{0}^{\epsilon}\left(-\ln y\right)^{-\alpha/2}\left[\frac{1}{\sqrt{y}}-2\sqrt{y}+\mathcal{O}\left(y^{\frac{3}{2}}\right)\right]dy.\label{eq:series}
\end{equation}
Noticing $\sqrt{(1-y)/[y(1+y)^{3}]}\leq\sqrt{1/y},$we find $I_{\alpha}(\epsilon)$
is bounded from below by the first term of its series expansion as

\begin{align}
I_{\alpha}(\epsilon) & \leq\int_{0}^{\epsilon}\left(-\ln y\right)^{-\alpha/2}\frac{1}{\sqrt{y}}dy\\
 & =2\sqrt{\epsilon}\left[\ln\left(\epsilon^{-1}\right)\right]^{-\frac{\alpha}{2}}-\alpha\int_{0}^{\epsilon}\left[\ln\left(y^{-1}\right)\right]^{-\frac{\alpha}{2}-1}y^{-\frac{1}{2}}dy\\
 & \leq2\sqrt{\epsilon}\left[\ln\left(\epsilon^{-1}\right)\right]^{-\frac{\alpha}{2}}.\label{eq:18}
\end{align}
Therefore, the thermodynamic length $\mathcal{L}(0)=\sqrt{\beta^{\left(\alpha-1\right)}\gamma_{0}^{-1}}f_{\alpha}(\epsilon)$
has a lower bound as

\begin{align}
\mathcal{L}(\epsilon) & =\sqrt{\beta^{(\alpha-1)}\gamma_{0}^{-1}}\left[f_{\alpha}(0)-I_{\alpha}(\epsilon)\right]\\
 & \geq\sqrt{\beta^{(\alpha-1)}\gamma_{0}^{-1}}\left[f_{\alpha}(0)-2\sqrt{\epsilon}\left(-\ln\epsilon\right)^{-\alpha/2}\right]\\
 & =\mathcal{L}(0)-2\sqrt{\beta^{\left(\alpha-1\right)}\gamma_{0}^{-1}\epsilon\ln^{-\alpha}\left(\epsilon^{-1}\right)}.\label{eq:Le>L0}
\end{align}
We stress that this inequality for $\mathcal{L}(\epsilon)$ is quite
tight in the low error probability regime ($\epsilon\ll1$). As shown
in Fig. \ref{fig:-as-the-1}, the asymptotic expression

\begin{equation}
\mathcal{L}(\epsilon)\approx\mathcal{L}(0)-2\sqrt{\beta^{\left(\alpha-1\right)}\gamma_{0}^{-1}\epsilon\ln^{-\alpha}\left(\epsilon^{-1}\right)}.\label{eq:L-scaling}
\end{equation}
of the thermodynamic length (lines) agrees well with the exact numerical
results (dots) obtained from Eq. (\ref{eq:TL}).

\begin{figure}
\centering{}\includegraphics[width=10cm]{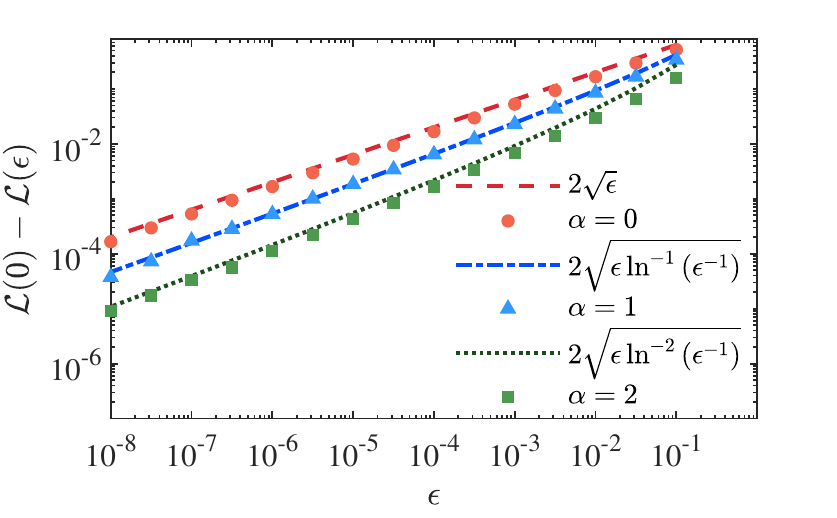}\caption{\label{fig:-as-the-1}$\mathcal{L}(0)-\mathcal{L}(\epsilon)$ as functions
of $\epsilon$ with $\alpha=0,1,2$. In this plot, we choose $\beta=1,\gamma_{0}=1$.}
\end{figure}
\end{widetext}

\begin{figure}
\begin{centering}
\subfloat[]{\includegraphics[width=9cm]{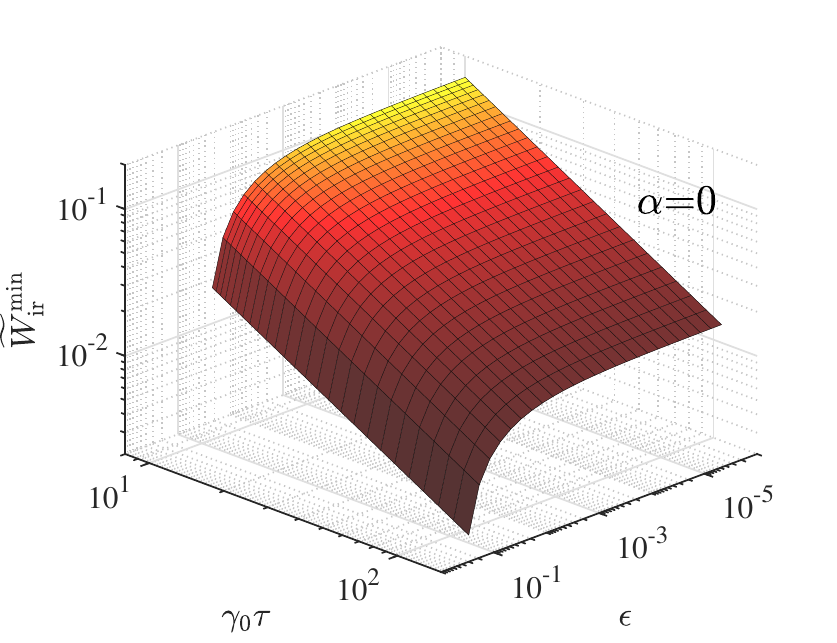}}\subfloat[]{\includegraphics[width=9cm]{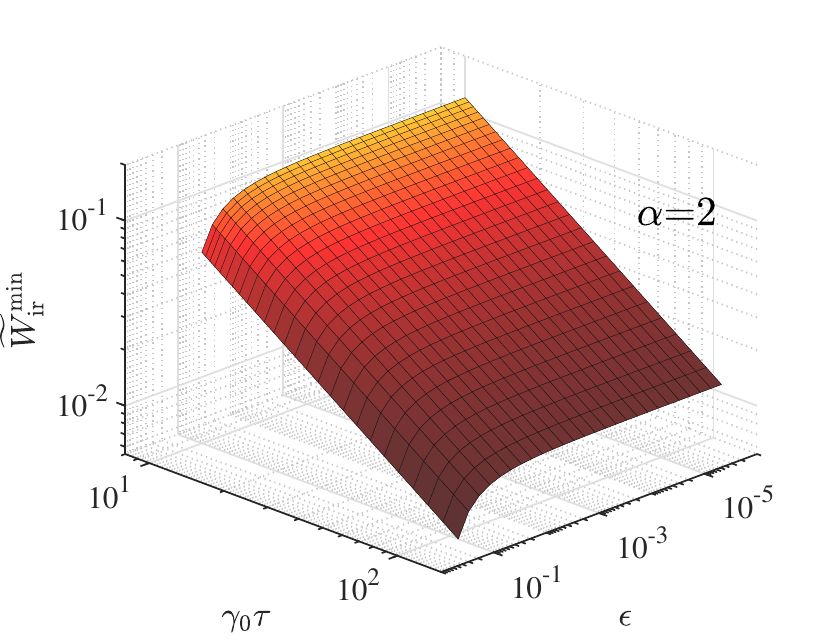}}
\par\end{centering}
\centering{}\caption{\label{fig:Analytical-trade-off-between-1} (a) Analytical trade-off
among irreversible work, erasure time and accuracy in the cases with
(a) $\alpha=0$ and (b) $\alpha=2$. The surface is plotted with $\widetilde{W}_{\mathrm{ir}}^{\mathrm{min}}(\epsilon)$
according to {[}Eq. (6){]} of the main text. In this plot, $\gamma_{0}=1$
and $\beta=1$ are used.}
\end{figure}

Substituting Eq. (\ref{eq:Le>L0}) into the bound on the irreversible
work $W_{\mathrm{ir}}\geq\mathcal{L}^{2}(\epsilon)/\tau$, one obtains

\begin{equation}
W_{\mathrm{ir}}\tau\geq\mathcal{L}^{2}(0)-4\mathcal{L}(0)\sqrt{\beta^{\left(\alpha-1\right)}\gamma_{0}^{-1}\epsilon\ln^{-\alpha}\left(\epsilon^{-1}\right)},
\end{equation}
where the $\mathcal{O}(\epsilon)$ term has been ignored. Then, divide
$\mathcal{L}^{2}(0)$ on both sides of the above inequality, we find

\begin{equation}
\frac{W_{\mathrm{ir}}\tau}{\mathcal{L}^{2}(0)}+\mu_{\alpha}\sqrt{\epsilon\ln^{-\alpha}\left(\epsilon^{-1}\right)}\geq1
\end{equation}
with $\mu_{\alpha}\equiv4/f_{\alpha}(0)$. This trade-off relation
among irreversible work, erasure time, and error probability is presented
as {[}Eq. (6){]} in the main text. The equality gives the analytical
lower bound $\widetilde{W}_{\mathrm{ir}}^{\mathrm{min}}(\epsilon)\equiv(1-\mu_{\alpha}\sqrt{\epsilon\ln^{-\alpha}\left(\epsilon^{-1}\right)})\mathcal{L}^{2}(0)/\tau$
for irreversible work. In Fig. \ref{fig:Analytical-trade-off-between-1},
the analytical lower bound ($\widetilde{W}_{\mathrm{ir}}^{\mathrm{min}}$)
of the trade-off in {[}Eq. (6){]} of the main text is plotted as the
surface for (a) $\alpha=0$ and (b) $\alpha=2$. 

\section{\label{sec:Optimal-control-for}Optimal erasure protocol to achieve
minimum irreversible work}
\begin{widetext}
\begin{figure}
\centering{}\subfloat[]{\includegraphics[width=9cm]{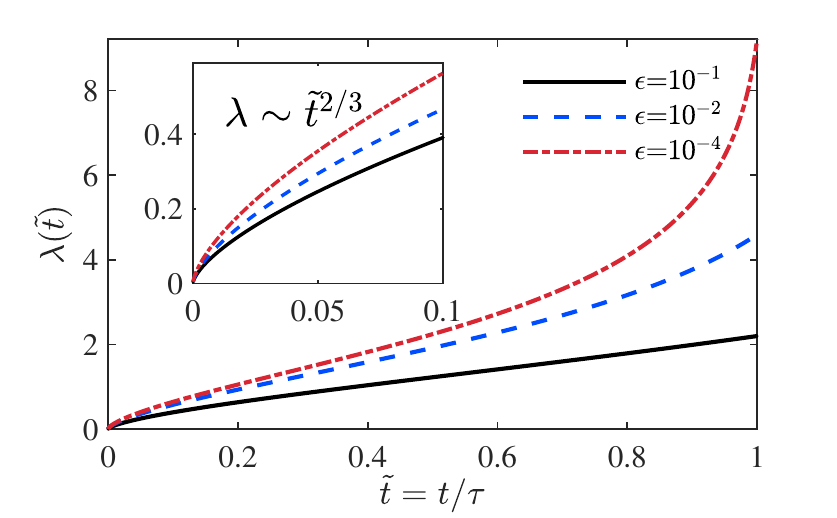}}\subfloat[]{\includegraphics[width=9cm]{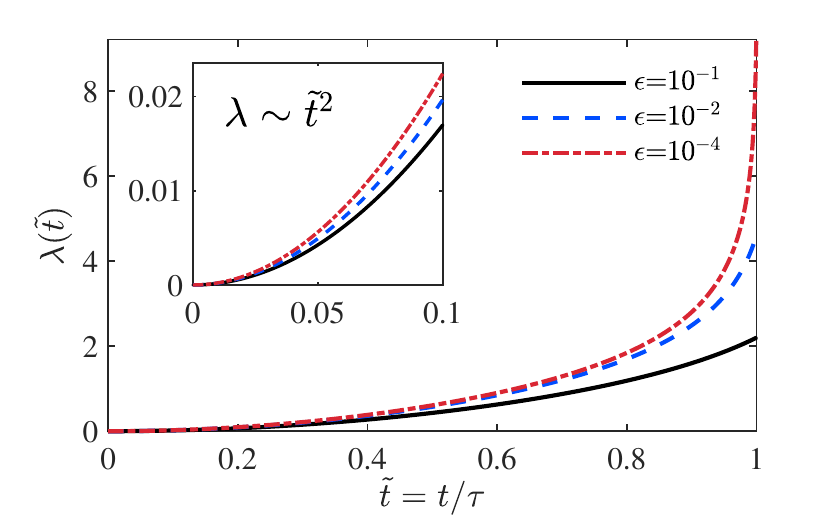}}\caption{\label{fig:Optimal-control-scheme-2-1}Optimal protocol of $\lambda(\tilde{t})$
in the cases with (a) $\alpha=0$ and (b) $\alpha=2$ for different
error probability $\epsilon$. The black solid curve, blue dashed
curve, and red dash-dotted curve represent the optimal protocols for
$\epsilon=10^{-1}$, $\epsilon=10^{-2}$, and $\epsilon=10^{-4}$,
respectively. The scaling of the optimal protocol in the initial stage
$\lambda(\widetilde{t}\ll1)\sim\tilde{t}^{2/(3-\alpha)}$ is illustrated
in the inset figure. In this plot, $\gamma_{0}=1$ and $\beta=1$
are used.}
\end{figure}
By numerically solving {[}Eq. (7){]} of the main text, the exact optimal
protocols for $\alpha=0,2$ are plotted in Fig. \ref{fig:Optimal-control-scheme-2-1}.
In the initial stage ($\tilde{t}=t/\tau\ll1$) of the population reduction
process, noticing $\lambda(\tilde{t})\ll1$, {[}Eq. (7){]} of the
main text is approximated as

\begin{equation}
\frac{d\lambda}{d\tilde{t}}\approx\mathcal{L}(\epsilon)\left[\frac{\beta\left(\beta\lambda\right)\left(1-\beta\lambda\right)}{\gamma_{0}\lambda^{\alpha}\left(2-\beta\lambda\right)^{3}}\right]^{-\frac{1}{2}}\approx\mathcal{L}(\epsilon)\beta^{-1}\sqrt{8\gamma_{0}}\lambda^{\frac{\alpha-1}{2}}.
\end{equation}
By straightforward calculation, one obtains the optimal protocol

\begin{equation}
\lambda(\tilde{t}\ll1)=\left[2(3-\alpha)^{2}\gamma_{0}\beta^{-2}\mathcal{L}^{2}(\epsilon)\right]^{\frac{1}{3-\alpha}}\tilde{t}^{\frac{2}{3-\alpha}},\label{eq:lambdat<<1}
\end{equation}
where the power exponent of $\widetilde{t}$ is only determined by
the bath spectral parameter $\alpha$. In the initial stage, the scaling
of the optimal protocol, $\lambda\sim\tilde{t}^{2/(3-\alpha)}$, is
shown in the inset figure of Fig. \ref{fig:Optimal-control-scheme-2-1}.
\end{widetext}

\bibliographystyle{apsrev}
\bibliography{mainref}